\documentclass[prl,twocolumn]{revtex4-1}

\usepackage{graphicx}
\usepackage{bm}

\begin{document}
\title{A molecular superfluid: non-classical rotations in doped para-hydrogen clusters}
\author{Hui Li,$^{1,2}$ Robert J. Le Roy,$^{1}$ Pierre-Nicholas Roy,$^{1,*}$ and A.R.W.\ McKellar$^{3,\dagger}$
}
\affiliation{$^1$ Department of Chemistry, University of Waterloo, Waterloo, Ontario~ N2L 3G1, Canada \\
$^2$ Institute of Theoretical Chemistry, State Key Laboratory of Theoretical
and Computational Chemistry, 
Jilin University, 2519 Jeifang Road, Changchun 130023, People's Republic of China\\
$^3$ Steacie Institute for Molecular Sciences, National Research Council of
Canada, Ottawa,
Ontario~ K1A 0R6, Canada \\
$^*$pnroy@uwaterloo.ca\\
$^\dagger$Robert.McKellar@nrc-cnrc.gc.ca
}
\date{\today; Published as: Phys. Rev. Lett. {\bf 105}, 133401 (2010)}

\begin{abstract}
Clusters of para-hydrogen ($p$H$_2$) have been predicted to exhibit superfluid behavior,
but direct observation of this phenomenon has been elusive. Combining
experiments and theoretical simulations, we have
determined the size evolution of the superfluid response of $p$H$_2$ clusters doped
with carbon dioxide (CO$_2$).  Reduction of the effective inertia is observed 
when the dopant is surrounded by the $p$H$_2$ solvent. 
This marks the onset of molecular superfluidity in $p$H$_2$. 
The fractional occupation of solvation rings around CO$_2$
correlates with enhanced superfluid response for certain cluster sizes.


\end{abstract}

\maketitle

Superfluidity has been well characterized in the bulk liquid phase \cite{Leggett1999}. In a breakthrough experiment, Grebenev, Toennies, and Vilesov observed superfluidity in helium nanodroplets \cite{Grebenev1998}, introducing a new way of investigating finite quantum systems \cite{Toennies2004}. Only atomic helium has been known to exhibit this behaviour at liquid-like densities, and the direct observation of superfluidity in a molecular system has remained elusive. The most likely candidate for a molecular superfluid is para-hydrogen ($p$H$_2$), because of its bosonic character, low mass, and weak intermolecular forces. In order to confirm superfluidity in molecular hydrogen, an experimental determination of the so-called superfluid response or non-classical rotational inertia \cite{Leggett1999} is required. 

The spectroscopic observation of nearly free molecular rotation in helium nanodroplets \cite{Grebenev1998, Toennies2004} has motivated several studies aimed at providing an understanding of superfluidity at the nanoscale in finite systems. Rovibrational spectra of linear molecules such as OCS in helium droplets are characterized by narrow lines  indicative of coherent rotation and decoupling from the solvent \cite{Grebenev1998}. Most importantly, the spectroscopic rotational 
constant $B$ \cite{Bnote} of the molecular dopant is ÒrenormalizedÓ by the helium environment. $B$ is inversely proportional to the effective moment of inertia of the system, so in a normal classical system, the inertia should grow monotonically with system size $N$. A key experiment on OCS-doped clusters with a few ($N = 1-8$) helium atoms has shown that their effective inertia could actually be greater than that of larger droplets \cite{Tang2002b}. 
Thus a decoupling mechanism must exist in order to stop the growth of the effective inertia as the cluster 
grows beyond $N = 8$. The turnaround marks the onset of superfluidity \cite{Tang2002b}. 

Theoretical analysis 
has
later shown that the size evolution of $B$ was related to that of the superfluid fraction, 
defined as the deviation from a classical response to rotation \cite{Paesani2005,Xu2006,Topic2006}. 
$B$ constants were also used to estimate the `experimental' superfluid 
fraction \cite{Xu2006,Topic2006}.  
Theorists have predicted superfluidity in $p$H$_2$ clusters \cite{Sindzingre1991},
but the direct observation of superfluid response in doped $p$H$_2$ clusters has remained elusive. 
Simulation work predicted superfluidity for OCS-doped $p$H$_2$ clusters \cite{Kwon2002,Paesani2005a} with a turnaround of $B$ at $N = 14$. 
However, this prediction is not yet confirmed by experiment, 
since published measurements only 
extend to $N = 7$ \cite{Michaud2008}. 
In other experiments, superfluidity was inferred for doped $p$H$_2$ clusters embedded in helium nanodroplets based on the behavior of the $Q$ branch of the 
rovibrational spectrum \cite{Grebenev2000c,Grebenev2002,Grebenev2008,Grebenev2010}.
Moore and Miller performed related experiments on doped HD 
clusters \cite{moore2003, moore2003a,moore2004}.
These results motivated our search for a system where the superfluidity of 
$p$H$_2$ could be discerned.

Here we report confirmation of molecular superfluid behavior in doped $p$H$_2$ clusters, based on spectroscopic determination of their non-classical rotational inertia probed by CO$_2$. Our main result confirms earlier theoretical predictions that $p$H$_2$ clusters could become superfluid \cite{Sindzingre1991}, and lends weight to experiments that have shown evidence of superfluidity in doped $p$H$_2$ clusters embedded in helium 
nanodroplets \cite{Grebenev2000c,Grebenev2002,Grebenev2008,Grebenev2010}.
We have measured infrared spectra and performed finite-temperature path-integral Monte Carlo (PIMC) simulations 
to support our findings. 

The experimental apparatus has been described elsewhere \cite{Brookes2004}.  Backing pressures in the range of 5 to 25 atmospheres were used, and the jet nozzle was cooled (-20 to  100 $^o$C). Effective cluster rotational temperatures were 0.3 to 1 K. Since the measurements were performed for $^{13}$C$^{16}$O$_2$ and $^{13}$C$^{18}$O$_2$, the vibrational frequency shifts and $B$ rotational parameters reported here were scaled to values appropriate to $^{12}$C$^{16}$O$_2$ for comparison with theory. Due to CO$_2$ nuclear spin statistics and the low temperature of the jet expansion, only three transitions are generally observed for 
each cluster: $R(0)$, $R(2)$, and $P(2)$. We fit these lines in terms of three parameters: band origin, $B$ 
rotational constant, $D$ centrifugal distortion constant. 

Details of the PIMC calculation method are described 
elsewhere \cite{Moroni2004,Blinov2005,Xu2006,Li2009}. 
It relies on the so-called worm algorithm \cite{Boninsegni2006a,Boninsegni2006b} to account for bosonic exchange. We used a recent H$_2$-H$_2$ interaction 
potential \cite{Patkowski2008} along with our own H$_2$-CO$_2$ potential energy surfaces \cite{Li2010}. The calculations were performed for a low temperature (0.5 K), with 256 translational and 128 rotational time slices.  
A total number of $10^7$-$10^8$ Monte Carlo steps for each simulation yielded satisfactory error bars.
\begin{figure}[h]
\begin{center}
\includegraphics[angle=0,width=\columnwidth]{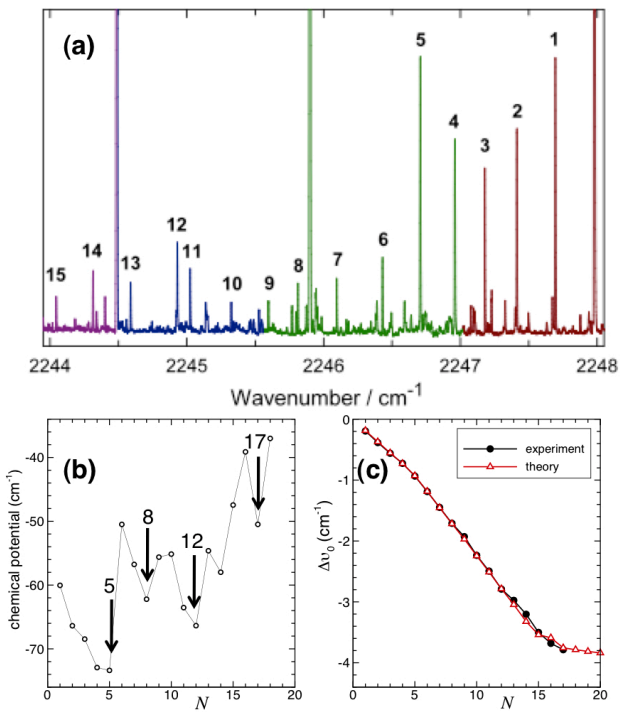}
\caption{(color online) (a) Observed spectrum 
with assigned $R(0)$ transitions labeled by cluster size, $N$. The three strongest (off-scale) transitions are the $P(4)$, $P(2)$, and $R(0)$ transitions of the CO$_2$ monomer. (b) Calculated chemical potential (in wavenumbers; 1.0 cm$^{-1}$ = 1.438769 K)  
versus $N$. (c) Measured (filled circles) and theoretical (open triangles) shifts of the CO$_2$ asymmetric stretch vibrational frequency versus $N$.}
\label{default}
\end{center}
\end{figure}
Figure 1a is a montage of 4 traces 
where the experimental conditions were optimized for each cluster size range. For this reason, and because laser power varies, relative line intensities are not quantitative. However, their qualitative trend is real, reflecting the behavior of the chemical potential (the energy difference between adjacent cluster sizes) curve in Fig. 1b, with $N =$ 5 and 12 being relatively stable (larger negative chemical potential) and $N =$ 6, 9 and 10 less so.
Line assignments for $N < 8$ are straightforward, and confirm the accuracy of the theoretical vibrational shifts. For $N \geq 8$ the line assignments are not so obvious. 
However, by using the theoretical CO$_2$ vibrational shifts in Fig. 1(c) to help guide the assignments, a convincing scheme is achieved which is more complete and self-consistent than any other. The special nature of $N=12$ is directly evident without any reference to theory: its transitions 
are noticeably stronger, indicating a local
 maximum in the cluster population, which is consistent with the local maximum in the absolute value of the chemical potential in Fig. 1(b).

Magic numbers indicative of particularly stable structures are observed for $N =$ 5, 8, 12 14, and 17.
In Fig. 1(c), the subtle change in slope at $N = 5$ is associated with filling of a `donut ring' of 5 $p$H$_2$ molecules around the CO$_2$ axis. For $5<N<9$ the additional $p$H$_2$ start populating a second ring.
The excellent agreement between the theoretical and experimental vibrational shifts attests to the quality of the interaction potentials used in the simulations.
\begin{figure}[h]
\begin{center}
\includegraphics[angle=0,width=\columnwidth]{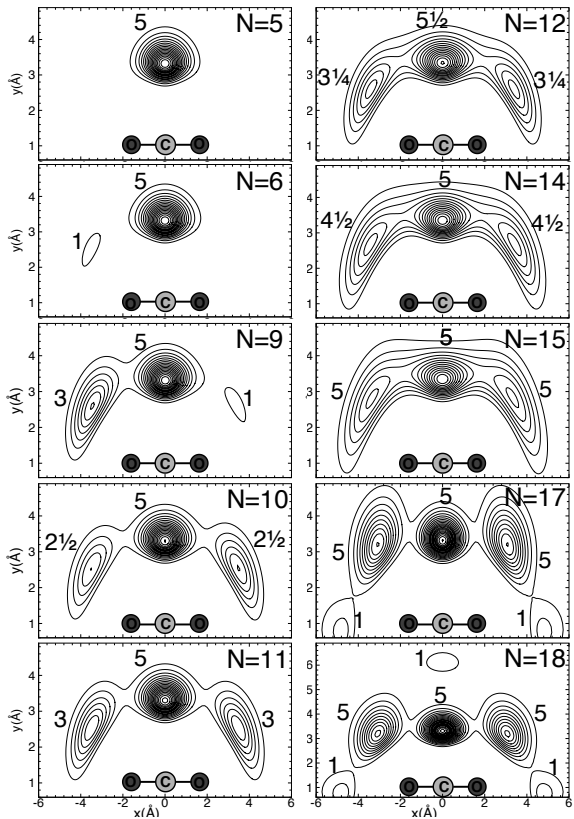}
\caption{ Contours of two dimensional projections of the $p$H$_2$ density in the CO$_2$ frame for representative $N$ values. Whole and fractional occupation numbers are shown. 
 }
\label{default}
\end{center}
\end{figure}

The structural evolution is shown in Fig. 2 by the patterns of the $p$H$_2$ density contours in the frame of the CO$_2$ molecule. 
Occupation numbers of `donut rings' and `terminal caps' are indicated. 
The occupation numbers, which add up to $N$, were obtained by integrating the $p$H$_2$ density in each ring and rounding 
to the nearest simple fraction. The boundary region between two rings was set as the minimum in the density. 
A central `donut ring' is filled for $N =$ 5 and a full solvation shell  is observed for $N = $17; the latter cluster size marks a more pronounced change in slope in the vibrational shift curve of Fig. 1c. The filling of a second centred ring starts at $N = 18$.

\begin{figure}[h]
\begin{center}
\includegraphics[angle=0,width=\columnwidth]{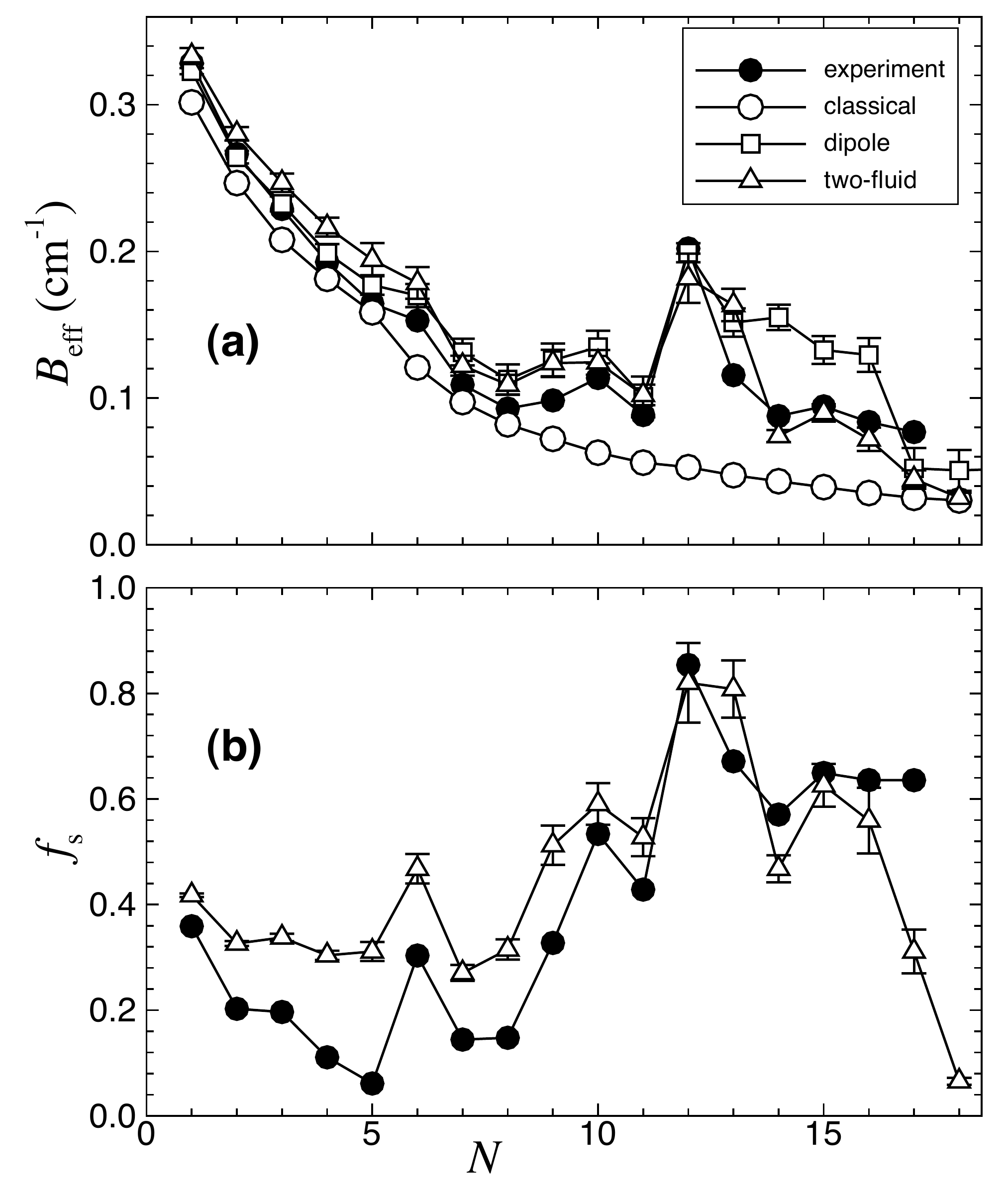}
\caption{(a) 
$B_{\rm eff}$ 
versus $N$ from experiment (filled circles), and dipole-dipole (open squares), two-fluid  (open triangles), and classical (open circles) calculations. (b) Experimental (filled circles) and calculated (open triangles) 
$f_s$ versus $N$.
}
\label{default}
\end{center}
\end{figure}

Figure 3a shows the effective $B$ constant, $B_{\rm eff}$ as a function $N$. Calculations based on 
a dipole-dipole correlation function can  be compared directly to experiment. In earlier work on helium clusters doped with CO$_2$  \cite{Blinov2005}, 
N$_2$O \cite{Xu2006,Moroni2004}, and HCCCN \cite{Topic2006}, the calculated $B$ was in very good agreement with experiment. The present theoretical values agree remarkably well with experiment, and capture the overall behavior of the size evolution of $B_{\rm eff}^{\rm exp}$. However, some deviations are observed for $N \geq 13$. These discrepancies may be attributed to the finite (non-zero) temperature of the simulations.

A key question is whether or not 
$B_{\rm eff}^{\rm exp}$
really probes superfluidity.  For systems undergoing rotation, the superfluid fraction is defined in terms of the non-classical rotational inertia using a two-fluid model \cite{Ceperley1995,Draeger2003} which assumes that the total density of the fluid 
is the sum of contributions from normal and superfluid components:  $\rho=\rho_s+\rho_n$. The superfluid fraction 
is   $f_s=\frac{\rho_s}{\rho} $ while the normal fraction is $ f_n=\frac{\rho_n}{\rho}$, defined as $f_n=\frac{I_{\rm eff}}{I_{\rm cl}} $, the ratio of the effective ($I_{\rm eff}$) over the classical ($I_{\rm cl}$) inertia  of $p$H$_2$. 
The calculated $f_s$ shown in Fig.3b corresponds
 to the perpendicular response with respect to the CO$_2$ axis. 
The experimental $f_s$ was obtained using
  the approach proposed in Ref. \cite{Xu2006}, according to which $B_{\rm eff}^{\rm exp}$ defines the total moment of inertia of the system via the relation  $B^{\rm exp}=\frac{\hbar^2}{2 I^{\rm exp}_{\rm total}}$. The effective inertia of $p$H$_2$ is obtained by subtracting the moment of inertia of CO$_2$, 
$I^{\rm exp}_{\rm eff}=I^{\rm exp}_{\rm total}-I_{\rm CO_2} $ . 
An experimental estimate of $f_s$ is presented in Fig. 3b.
The experimental and theoretical $f_s$ values agree fairly well in their 
overall behavior over a large range of sizes. 

We conclude that within linear response theory, CO$_2$ can be used as a probe of $f_s$ in $p$H$_2$ clusters. 
It is also possible to use the computed $f_s$ to obtain two-fluid $B$ constants. 
Such two-fluid values are presented  in Fig. 3a. Their behavior is very close to that of 
$B_{\rm eff}^{\rm exp}$, and further confirms that CO$_2$ is a probe of superfluid response. 
To highlight the non-classical behaviour of $I_{\rm eff}$, the effective $B$ constants based on classical cluster inertia ($B_{\rm cl}=\frac{\hbar^2}{2(I_{\rm cl}+I_{\rm CO_2})} $) are also shown in Fig. 3a. As expected, $B_{\rm cl}$ decreases monotonically with $N$. 

\begin{figure}[h]
\begin{center}
\includegraphics[angle=0,width=\columnwidth]{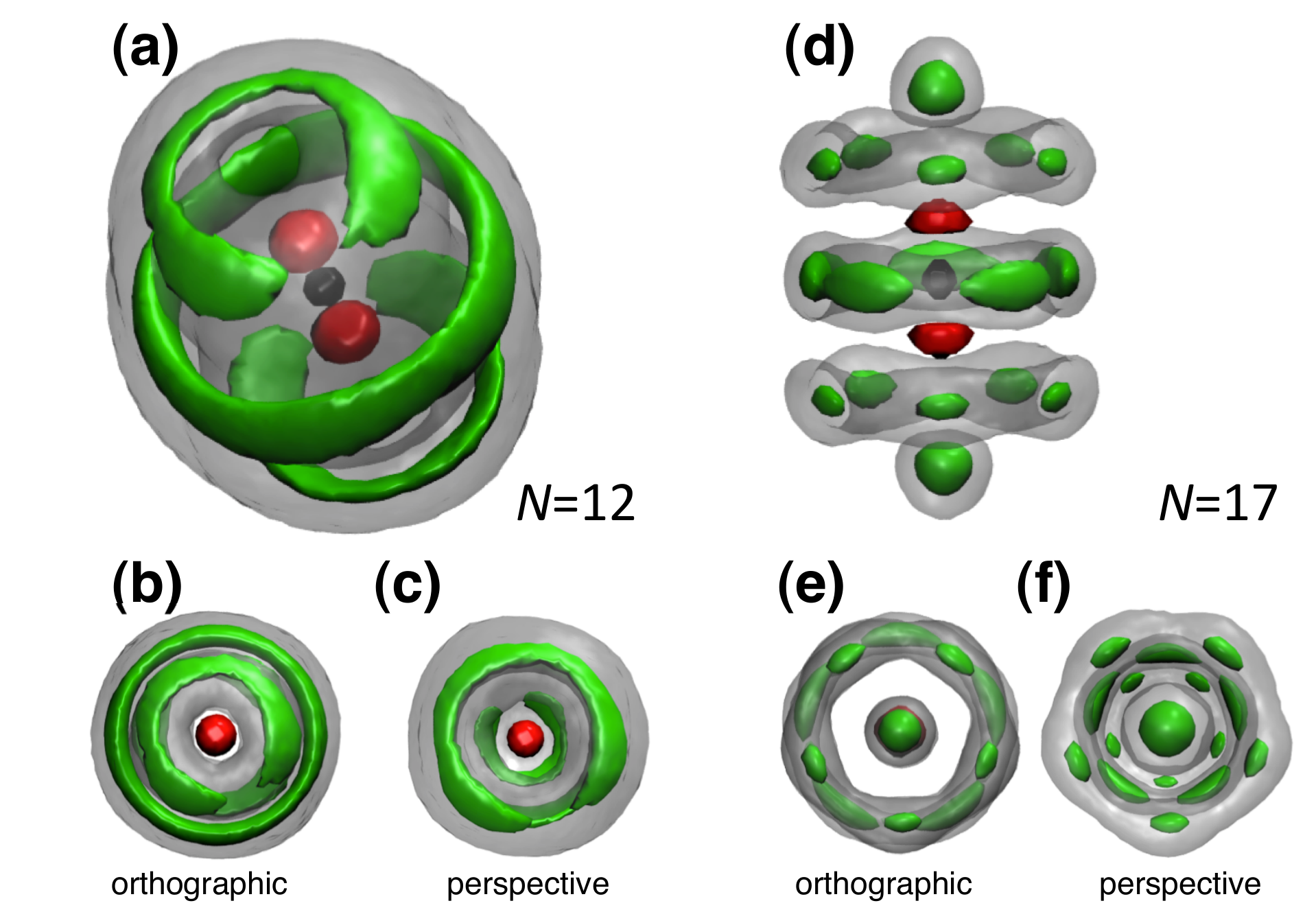}
\caption{(color online) Three-dimensional $p$H$_2$ densities \cite{framenote}. (a) Side view for $N=12$. (b) Top orthographic (or ÔflatÕ) view for $N=12$. (c) Perspective view for $N=12$. (d) side view for $N=17$. (e) Top orthographic view for $N=17$. (f) Perspective view for $N=17$. 
The green (dark) and grey (light) colors respectively represent high and low densities. 
}
\label{default}
\end{center}
\end{figure}

Figure 4 shows the low and high $p$H$_2$ density regions in 3D for two cluster sizes \cite{framenote}. 
For $N = 12$, the maximum in $f_s$, the low-density surface is continuous, except at the ends of the CO$_2$ axis. This is consistent with the enhanced bosonic exchanges associated with its larger $f_s$. 
For $N = 17$, the low-density ring regions are disconnected from one another, consistent with a lower $f_s$. For $N = 12$, the high-density surface is still highly delocalized within each ring, although the rings are no longer connected. For $N = 17$, pronounced localization is observed at high density, and one can clearly enumerate the individual $p$H$_2$ molecules. The orthographic top view for $N = 12$ (Fig. 4b) shows that the central ring is swollen to accommodate an additional fraction of a particle, while a similar view for $N = 17$ (Fig. 4e) shows that all three rings have similar radii with staggered particle placement.

Using CO$_2$ to probe superfluidity for doped $p$H$_2$ clusters in the $N=1-18$ range, we provide the first direct measurement of $f_s$ in a molecular system. A maximum in $f_s$ is observed at $N=12$. This value is confirmed by consistent experimental and theoretical results, and is associated with the appearance of fractional ring occupation. Beyond this size, $f_s$ decreases due to enhanced localization. 
There is a fine balance between the forces that cause localization, and bosonic exchanges that favour superfluididty.  Localization dominates at higher $N$.
A full solvation shell with pronounced localization is observed at $N = 17$; the computed $f_s$ there is substantially lower than the experimental value. A possible explanation for this discrepancy is the finite (non-zero) temperature character of the PIMC simulation and the fact this $f_s$ estimate is based on linear response theory. A lower temperature simulation could lead to a higher $f_s$ due to quantum 
melting \cite{Mezzacapo2006,Mezzacapo2007,Cuervo2008}. 
Whether or not larger clusters will be superfluid remains an open question. 
The localization modulated superfluid response may be viewed as the nanoscale analogue of the insulating to superfluid transitions observed in cold gases \cite{Greiner2002}.

\acknowledgments
We acknowledge N. Blinov, M. Gingras, W. J\"ager, F. McCourt, R. Melko, M. Nooijen, and P. Raston for discussions, the Natural Sciences and Engineering Research Council of Canada and the National Research Council of Canada for support, and the Shared Hierarchical Academic Research Computing Network for computing time.


\begin{thebibliography}{10}%
\makeatletter
\providecommand \@ifxundefined [1]{%
 \ifx #1\undefined \expandafter \@firstoftwo
 \else \expandafter \@secondoftwo
\fi
}%
\providecommand \@ifnum [1]{%
 \ifnum #1\expandafter \@firstoftwo
 \else \expandafter \@secondoftwo
\fi
}%
\providecommand \enquote [1]{``#1''}%
\providecommand \bibnamefont  [1]{#1}%
\providecommand \bibfnamefont [1]{#1}%
\providecommand \citenamefont [1]{#1}%
\providecommand\href[0]{\@sanitize\@href}%
\providecommand\@href[1]{\endgroup\@@startlink{#1}\endgroup\@@href}%
\providecommand\@@href[1]{#1\@@endlink}%
\providecommand \@sanitize [0]{\begingroup\catcode`\&12\catcode`\#12\relax}%
\@ifxundefined \pdfoutput {\@firstoftwo}{%
 \@ifnum{\z@=\pdfoutput}{\@firstoftwo}{\@secondoftwo}%
}{%
 \providecommand\@@startlink[1]{\leavevmode}%
 \providecommand\@@endlink[0]{}%
}{%
 \providecommand\@@startlink[1]{%
  \leavevmode
  \pdfstartlink
   attr{/Border[0 0 1 ]/H/I/C[0 1 1]}%
   user{/Subtype/Link/A<</Type/Action/S/URI/URI(#1)>>}%
  \relax
 }%
 \providecommand\@@endlink[0]{\pdfendlink}%
}%
\providecommand \url  [0]{\begingroup\@sanitize \@url }%
\providecommand \@url [1]{\endgroup\@href {#1}{\urlprefix}}%
\providecommand \urlprefix [0]{URL }%
\providecommand \Eprint[0]{\href }%
\@ifxundefined \urlstyle {%
  \providecommand \doi [1]{doi:\discretionary{}{}{}#1}%
}{%
  \providecommand \doi [0]{doi:\discretionary{}{}{}\begingroup
  \urlstyle{rm}\Url }%
}%
\providecommand \doibase [0]{http://dx.doi.org/}%
\providecommand \Doi[1]{\href{\doibase#1}}%
\providecommand \bibAnnote [3]{%
  \BibitemShut{#1}%
  \begin{quotation}\noindent
    \textsc{Key:}\ #2\\\textsc{Annotation:}\ #3%
  \end{quotation}%
}%
\providecommand \bibAnnoteFile [2]{%
  \IfFileExists{#2}{\bibAnnote {#1} {#2} {\input{#2}}}{}%
}%
\providecommand \typeout [0]{\immediate \write \m@ne }%
\providecommand \selectlanguage [0]{\@gobble}%
\providecommand \bibinfo [0]{\@secondoftwo}%
\providecommand \bibfield [0]{\@secondoftwo}%
\providecommand \translation [1]{[#1]}%
\providecommand \BibitemOpen[0]{}%
\providecommand \bibitemStop [0]{}%
\providecommand \bibitemNoStop [0]{.\EOS\space}%
\providecommand \EOS [0]{\spacefactor3000\relax}%
\providecommand \BibitemShut [1]{\csname bibitem#1\endcsname}%
\bibitem{Leggett1999}%
  \BibitemOpen
  \bibfield{author}{%
  \bibinfo {author} {\bibfnamefont{A.~J.}\ \bibnamefont{Leggett}},\ }%
  \bibfield{journal}{%
  \bibinfo {journal} {Rev. Mod. Phys.}\ }%
  \textbf{\bibinfo {volume} {71}},\ \bibinfo {pages} {S318} (\bibinfo {year}
  {1999})%
  \bibAnnoteFile{NoStop}{Leggett1999}%
\bibitem{Grebenev1998}%
  \BibitemOpen
  \bibfield{author}{%
  \bibinfo {author} {\bibfnamefont{S.}~\bibnamefont{Grebenev}}, \bibinfo
  {author} {\bibfnamefont{J.~P.}\ \bibnamefont{Toennies}},\ and\ \bibinfo
  {author} {\bibfnamefont{A.~F.}\ \bibnamefont{Vilesov}},\ }%
  \bibfield{journal}{%
  \bibinfo {journal} {Science}\ }%
  \textbf{\bibinfo {volume} {279}},\ \bibinfo {pages} {2083} (\bibinfo {year}
  {1998})%
  \bibAnnoteFile{NoStop}{Grebenev1998}%
\bibitem{Toennies2004}%
  \BibitemOpen
  \bibfield{author}{%
  \bibinfo {author} {\bibfnamefont{J.~P.}\ \bibnamefont{Toennies}}\ and\
  \bibinfo {author} {\bibfnamefont{A.~F.}\ \bibnamefont{Vilesov}},\ }%
  \bibfield{journal}{%
  \bibinfo {journal} {Angew. Chem.-Int. Edit.}\ }%
  \textbf{\bibinfo {volume} {43}},\ \bibinfo {pages} {2622} (\bibinfo {year}
  {2004})%
  \bibAnnoteFile{NoStop}{Toennies2004}%
\bibitem{Bnote}%
  \BibitemOpen
  \bibinfo {note} {Half the energy spacing between the ground and first excited
  rotational level.}%
  \bibAnnoteFile{Stop}{Bnote}%
\bibitem{Tang2002b}%
  \BibitemOpen
  \bibfield{author}{%
  \bibinfo {author} {\bibfnamefont{J.}~\bibnamefont{Tang}}, \bibinfo {author}
  {\bibfnamefont{Y.~J.}\ \bibnamefont{Xu}}, \bibinfo {author}
  {\bibfnamefont{A.~R.~W.}\ \bibnamefont{McKellar}},\ and\ \bibinfo {author}
  {\bibfnamefont{W.}~\bibnamefont{J\"ager}},\ }%
  \bibfield{journal}{%
  \bibinfo {journal} {Science}\ }%
  \textbf{\bibinfo {volume} {297}},\ \bibinfo {pages} {2030} (\bibinfo {year}
  {2002})%
  \bibAnnoteFile{NoStop}{Tang2002b}%
\bibitem{Paesani2005}%
  \BibitemOpen
  \bibfield{author}{%
  \bibinfo {author} {\bibfnamefont{F.}~\bibnamefont{Paesani}}, \bibinfo
  {author} {\bibfnamefont{Y.}~\bibnamefont{Kwon}},\ and\ \bibinfo {author}
  {\bibfnamefont{K.~B.}\ \bibnamefont{Whaley}},\ }%
  \bibfield{journal}{%
  \bibinfo {journal} {Phys. Rev. Lett.}\ }%
  \textbf{\bibinfo {volume} {94}},\ \bibinfo {pages} {153401} (\bibinfo {year}
  {2005})%
  \bibAnnoteFile{NoStop}{Paesani2005}%
\bibitem{Xu2006}%
  \BibitemOpen
  \bibfield{author}{%
  \bibinfo {author} {\bibfnamefont{Y.~J.}\ \bibnamefont{Xu}}, \bibinfo {author}
  {\bibfnamefont{N.}~\bibnamefont{Blinov}}, \bibinfo {author}
  {\bibfnamefont{W.}~\bibnamefont{J\"ager}},\ and\ \bibinfo {author}
  {\bibfnamefont{P.-N.}\ \bibnamefont{Roy}},\ }%
  \bibfield{journal}{%
  \bibinfo {journal} {J. Chem. Phys.}\ }%
  \textbf{\bibinfo {volume} {124}},\ \bibinfo {pages} {081101} (\bibinfo {year}
  {2006})%
  \bibAnnoteFile{NoStop}{Xu2006}%
\bibitem{Topic2006}%
  \BibitemOpen
  \bibfield{author}{%
  \bibinfo {author} {\bibfnamefont{W.}~\bibnamefont{Topic}}, \bibinfo {author}
  {\bibfnamefont{W.}~\bibnamefont{J\"ager}}, \bibinfo {author}
  {\bibfnamefont{N.}~\bibnamefont{Blinov}}, \bibinfo {author}
  {\bibfnamefont{P.-N.}\ \bibnamefont{Roy}}, \bibinfo {author}
  {\bibfnamefont{M.}~\bibnamefont{Botti}},\ and\ \bibinfo {author}
  {\bibfnamefont{S.}~\bibnamefont{Moroni}},\ }%
  \bibfield{journal}{%
  \bibinfo {journal} {J. Chem. Phys.}\ }%
  \textbf{\bibinfo {volume} {125}},\ \bibinfo {pages} {144310} (\bibinfo {year}
  {2006})%
  \bibAnnoteFile{NoStop}{Topic2006}%
\bibitem{Sindzingre1991}%
  \BibitemOpen
  \bibfield{author}{%
  \bibinfo {author} {\bibfnamefont{P.}~\bibnamefont{Sindzingre}}, \bibinfo
  {author} {\bibfnamefont{D.~M.}\ \bibnamefont{Ceperley}},\ and\ \bibinfo
  {author} {\bibfnamefont{M.~L.}\ \bibnamefont{Klein}},\ }%
  \bibfield{journal}{%
  \bibinfo {journal} {Phys. Rev. Lett.}\ }%
  \textbf{\bibinfo {volume} {67}},\ \bibinfo {pages} {1871} (\bibinfo {year}
  {1991})%
  \bibAnnoteFile{NoStop}{Sindzingre1991}%
\bibitem{Kwon2002}%
  \BibitemOpen
  \bibfield{author}{%
  \bibinfo {author} {\bibfnamefont{Y.}~\bibnamefont{Kwon}}\ and\ \bibinfo
  {author} {\bibfnamefont{K.~B.}\ \bibnamefont{Whaley}},\ }%
  \bibfield{journal}{%
  \bibinfo {journal} {Phys. Rev. Lett.}\ }%
  \textbf{\bibinfo {volume} {89}},\ \bibinfo {pages} {273401} (\bibinfo {year}
  {2002})%
  \bibAnnoteFile{NoStop}{Kwon2002}%
\bibitem{Paesani2005a}%
  \BibitemOpen
  \bibfield{author}{%
  \bibinfo {author} {\bibfnamefont{F.}~\bibnamefont{Paesani}}, \bibinfo
  {author} {\bibfnamefont{R.~E.}\ \bibnamefont{Zillich}}, \bibinfo {author}
  {\bibfnamefont{Y.}~\bibnamefont{Kwon}},\ and\ \bibinfo {author}
  {\bibfnamefont{K.~B.}\ \bibnamefont{Whaley}},\ }%
  \bibfield{journal}{%
  \bibinfo {journal} {J. Chem. Phys.}\ }%
  \textbf{\bibinfo {volume} {122}},\ \bibinfo {pages} {181106} (\bibinfo {year}
  {2005})%
  \bibAnnoteFile{NoStop}{Paesani2005a}%
\bibitem{Michaud2008}%
  \BibitemOpen
  \bibfield{author}{%
  \bibinfo {author} {\bibfnamefont{J.~M.}\ \bibnamefont{Michaud}}\ and\
  \bibinfo {author} {\bibfnamefont{W.}~\bibnamefont{J\"ager}},\ }%
  \bibfield{journal}{%
  \bibinfo {journal} {J. Chem. Phys.}\ }%
  \textbf{\bibinfo {volume} {129}},\ \bibinfo {pages} {144311} (\bibinfo {year}
  {2008})%
  \bibAnnoteFile{NoStop}{Michaud2008}%
\bibitem{Grebenev2000c}%
  \BibitemOpen
  \bibfield{author}{%
  \bibinfo {author} {\bibfnamefont{S.}~\bibnamefont{Grebenev}}, \bibinfo
  {author} {\bibfnamefont{B.}~\bibnamefont{Sartakov}}, \bibinfo {author}
  {\bibfnamefont{J.~P.}\ \bibnamefont{Toennies}},\ and\ \bibinfo {author}
  {\bibfnamefont{A.~F.}\ \bibnamefont{Vilesov}},\ }%
  \bibfield{journal}{%
  \bibinfo {journal} {Science}\ }%
  \textbf{\bibinfo {volume} {289}},\ \bibinfo {pages} {1532} (\bibinfo {year}
  {2000})%
  \bibAnnoteFile{NoStop}{Grebenev2000c}%
\bibitem{Grebenev2002}%
  \BibitemOpen
  \bibfield{author}{%
  \bibinfo {author} {\bibfnamefont{S.}~\bibnamefont{Grebenev}}, \bibinfo
  {author} {\bibfnamefont{B.}~\bibnamefont{Sartakov}}, \bibinfo {author}
  {\bibfnamefont{J.~P.}\ \bibnamefont{Toennies}},\ and\ \bibinfo {author}
  {\bibfnamefont{A.}~\bibnamefont{Vilesov}},\ }%
  \bibfield{journal}{%
  \bibinfo {journal} {Phys. Rev. Lett.}\ }%
  \textbf{\bibinfo {volume} {89}},\ \bibinfo {pages} {225301} (\bibinfo {year}
  {2002})%
  \bibAnnoteFile{NoStop}{Grebenev2002}%
\bibitem{Grebenev2008}%
  \BibitemOpen
  \bibfield{author}{%
  \bibinfo {author} {\bibfnamefont{S.}~\bibnamefont{Grebenev}}, \bibinfo
  {author} {\bibfnamefont{B.~G.}\ \bibnamefont{Sartakov}}, \bibinfo {author}
  {\bibfnamefont{J.~P.}\ \bibnamefont{Toennies}},\ and\ \bibinfo {author}
  {\bibfnamefont{A.~F.}\ \bibnamefont{Vilesov}},\ }%
  \bibfield{journal}{%
  \bibinfo {journal} {Epl-Europhys Lett}\ }%
  \textbf{\bibinfo {volume} {83}},\ \bibinfo {pages} {66008} (\bibinfo {year}
  {2008})%
  \bibAnnoteFile{NoStop}{Grebenev2008}%
\bibitem{Grebenev2010}%
  \BibitemOpen
  \bibfield{author}{%
  \bibinfo {author} {\bibfnamefont{S.}~\bibnamefont{Grebenev}}, \bibinfo
  {author} {\bibfnamefont{B.~G.}\ \bibnamefont{Sartakov}}, \bibinfo {author}
  {\bibfnamefont{J.~P.}\ \bibnamefont{Toennies}},\ and\ \bibinfo {author}
  {\bibfnamefont{A.~F.}\ \bibnamefont{Vilesov}},\ }%
  \bibfield{journal}{%
  \bibinfo {journal} {J. Chem. Phys.}\ }%
  \textbf{\bibinfo {volume} {132}},\ \bibinfo {pages} {064501} (\bibinfo {year}
  {2010})%
  \bibAnnoteFile{NoStop}{Grebenev2010}%
\bibitem{moore2003}%
  \BibitemOpen
  \bibfield{author}{%
  \bibinfo {author} {\bibfnamefont{D.~T.}\ \bibnamefont{Moore}}\ and\ \bibinfo
  {author} {\bibfnamefont{R.~E.}\ \bibnamefont{Miller}},\ }%
  \bibfield{journal}{%
  \bibinfo {journal} {J. Chem. Phys.}\ }%
  \textbf{\bibinfo {volume} {119}},\ \bibinfo {pages} {4713} (\bibinfo {year}
  {2003})%
  \bibAnnoteFile{NoStop}{moore2003}%
\bibitem{moore2003a}%
  \BibitemOpen
  \bibfield{author}{%
  \bibinfo {author} {\bibfnamefont{D.~T.}\ \bibnamefont{Moore}}\ and\ \bibinfo
  {author} {\bibfnamefont{R.~E.}\ \bibnamefont{Miller}},\ }%
  \bibfield{journal}{%
  \bibinfo {journal} {J. Phys. Chem. A}\ }%
  \textbf{\bibinfo {volume} {107}},\ \bibinfo {pages} {10805} (\bibinfo {year}
  {2003})%
  \bibAnnoteFile{NoStop}{moore2003a}%
\bibitem{moore2004}%
  \BibitemOpen
  \bibfield{author}{%
  \bibinfo {author} {\bibfnamefont{D.~T.}\ \bibnamefont{Moore}}\ and\ \bibinfo
  {author} {\bibfnamefont{R.~E.}\ \bibnamefont{Miller}},\ }%
  \bibfield{journal}{%
  \bibinfo {journal} {J. Phys. Chem. A}\ }%
  \textbf{\bibinfo {volume} {108}},\ \bibinfo {pages} {1930} (\bibinfo {year}
  {2004})%
  \bibAnnoteFile{NoStop}{moore2004}%
\bibitem{Brookes2004}%
  \BibitemOpen
  \bibfield{author}{%
  \bibinfo {author} {\bibfnamefont{M.~D.}\ \bibnamefont{Brookes}}, \bibinfo
  {author} {\bibfnamefont{C.~H.}\ \bibnamefont{Xia}}, \bibinfo {author}
  {\bibfnamefont{J.}~\bibnamefont{Tang}}, \bibinfo {author}
  {\bibfnamefont{J.~A.}\ \bibnamefont{Anstey}}, \bibinfo {author}
  {\bibfnamefont{B.~G.}\ \bibnamefont{Fulsom}}, \bibinfo {author}
  {\bibfnamefont{K.~X.~A.}\ \bibnamefont{Yong}}, \bibinfo {author}
  {\bibfnamefont{J.~M.}\ \bibnamefont{King}},\ and\ \bibinfo {author}
  {\bibfnamefont{A.~R.~W.}\ \bibnamefont{McKellar}},\ }%
  \bibfield{journal}{%
  \bibinfo {journal} {Spectrochim. Acta A}\ }%
  \textbf{\bibinfo {volume} {60}},\ \bibinfo {pages} {3235} (\bibinfo {year}
  {2004})%
  \bibAnnoteFile{NoStop}{Brookes2004}%
\bibitem{Moroni2004}%
  \BibitemOpen
  \bibfield{author}{%
  \bibinfo {author} {\bibfnamefont{S.}~\bibnamefont{Moroni}}, \bibinfo {author}
  {\bibfnamefont{N.}~\bibnamefont{Blinov}},\ and\ \bibinfo {author}
  {\bibfnamefont{P.-N.}\ \bibnamefont{Roy}},\ }%
  \bibfield{journal}{%
  \bibinfo {journal} {J. Chem. Phys.}\ }%
  \textbf{\bibinfo {volume} {121}},\ \bibinfo {pages} {3577} (\bibinfo {year}
  {2004})%
  \bibAnnoteFile{NoStop}{Moroni2004}%
\bibitem{Blinov2005}%
  \BibitemOpen
  \bibfield{author}{%
  \bibinfo {author} {\bibfnamefont{N.}~\bibnamefont{Blinov}}\ and\ \bibinfo
  {author} {\bibfnamefont{P.-N.}\ \bibnamefont{Roy}},\ }%
  \bibfield{journal}{%
  \bibinfo {journal} {J. Low. Temp. Phys.}\ }%
  \textbf{\bibinfo {volume} {140}},\ \bibinfo {pages} {253} (\bibinfo {year}
  {2005})%
  \bibAnnoteFile{NoStop}{Blinov2005}%
\bibitem{Li2009}%
  \BibitemOpen
  \bibfield{author}{%
  \bibinfo {author} {\bibfnamefont{H.}~\bibnamefont{Li}}, \bibinfo {author}
  {\bibfnamefont{N.}~\bibnamefont{Blinov}}, \bibinfo {author}
  {\bibfnamefont{P.-N.}\ \bibnamefont{Roy}},\ and\ \bibinfo {author}
  {\bibfnamefont{R.~J.}\ \bibnamefont{Le~Roy}},\ }%
  \bibfield{journal}{%
  \bibinfo {journal} {J. Chem. Phys.}\ }%
  \textbf{\bibinfo {volume} {130}},\ \bibinfo {pages} {144305} (\bibinfo {year}
  {2009})%
  \bibAnnoteFile{NoStop}{Li2009}%
\bibitem{Boninsegni2006a}%
  \BibitemOpen
  \bibfield{author}{%
  \bibinfo {author} {\bibfnamefont{M.}~\bibnamefont{Boninsegni}}, \bibinfo
  {author} {\bibfnamefont{N.~V.}\ \bibnamefont{Prokof'ev}},\ and\ \bibinfo
  {author} {\bibfnamefont{B.~V.}\ \bibnamefont{Svistunov}},\ }%
  \bibfield{journal}{%
  \bibinfo {journal} {Phys. Rev. Lett.}\ }%
  \textbf{\bibinfo {volume} {96}},\ \bibinfo {pages} {070601} (\bibinfo {year}
  {2006})%
  \bibAnnoteFile{NoStop}{Boninsegni2006a}%
\bibitem{Boninsegni2006b}%
  \BibitemOpen
  \bibfield{author}{%
  \bibinfo {author} {\bibfnamefont{M.}~\bibnamefont{Boninsegni}}, \bibinfo
  {author} {\bibfnamefont{N.~V.}\ \bibnamefont{Prokof'ev}},\ and\ \bibinfo
  {author} {\bibfnamefont{B.~V.}\ \bibnamefont{Svistunov}},\ }%
  \bibfield{journal}{%
  \bibinfo {journal} {Phys. Rev. E}\ }%
  \textbf{\bibinfo {volume} {74}},\ \bibinfo {pages} {036701} (\bibinfo {year}
  {2006})%
  \bibAnnoteFile{NoStop}{Boninsegni2006b}%
\bibitem{Patkowski2008}%
  \BibitemOpen
  \bibfield{author}{%
  \bibinfo {author} {\bibfnamefont{K.}~\bibnamefont{Patkowski}}, \bibinfo
  {author} {\bibfnamefont{W.}~\bibnamefont{Cencek}}, \bibinfo {author}
  {\bibfnamefont{P.}~\bibnamefont{Jankowski}}, \bibinfo {author}
  {\bibfnamefont{K.}~\bibnamefont{Szalewicz}}, \bibinfo {author}
  {\bibfnamefont{J.~B.}\ \bibnamefont{Mehl}}, \bibinfo {author}
  {\bibfnamefont{G.}~\bibnamefont{Garberoglio}},\ and\ \bibinfo {author}
  {\bibfnamefont{A.~H.}\ \bibnamefont{Harvey}},\ }%
  \bibfield{journal}{%
  \bibinfo {journal} {J. Chem. Phys.}\ }%
  \textbf{\bibinfo {volume} {129}},\ \bibinfo {pages} {094304} (\bibinfo {year}
  {2008})%
  \bibAnnoteFile{NoStop}{Patkowski2008}%
\bibitem{Li2010}%
  \BibitemOpen
  \bibfield{author}{%
  \bibinfo {author} {\bibfnamefont{H.}~\bibnamefont{Li}}, \bibinfo {author}
  {\bibfnamefont{P.-N.}\ \bibnamefont{Roy}},\ and\ \bibinfo {author}
  {\bibfnamefont{R.~J.}\ \bibnamefont{Le~Roy}},\ }%
  \bibfield{journal}{%
  \bibinfo {journal} {J. Chem. Phys.}\ }%
  \textbf{\bibinfo {volume} {132}},\ \bibinfo {pages} {214309} (\bibinfo {year}
  {2010})%
  \bibAnnoteFile{NoStop}{Li2010}%
\bibitem{Ceperley1995}%
  \BibitemOpen
  \bibfield{author}{%
  \bibinfo {author} {\bibfnamefont{D.~M.}\ \bibnamefont{Ceperley}},\ }%
  \bibfield{journal}{%
  \bibinfo {journal} {Rev. Mod. Phys.}\ }%
  \textbf{\bibinfo {volume} {67}},\ \bibinfo {pages} {279} (\bibinfo {year}
  {1995})%
  \bibAnnoteFile{NoStop}{Ceperley1995}%
\bibitem{Draeger2003}%
  \BibitemOpen
  \bibfield{author}{%
  \bibinfo {author} {\bibfnamefont{E.~W.}\ \bibnamefont{Draeger}}\ and\
  \bibinfo {author} {\bibfnamefont{D.~M.}\ \bibnamefont{Ceperley}},\ }%
  \bibfield{journal}{%
  \bibinfo {journal} {Phys. Rev. Lett.}\ }%
  \textbf{\bibinfo {volume} {90}},\ \bibinfo {pages} {065301} (\bibinfo {year}
  {2003})%
  \bibAnnoteFile{NoStop}{Draeger2003}%
\bibitem{framenote}%
  \BibitemOpen
  \bibinfo {note} {The densities were calculated in a body-fixed frame
  determined from the instantaneous moment of inertia tensor of the whole
  cluster at each Monte Carlo step.}%
  \bibAnnoteFile{Stop}{framenote}%
\bibitem{Mezzacapo2006}%
  \BibitemOpen
  \bibfield{author}{%
  \bibinfo {author} {\bibfnamefont{F.}~\bibnamefont{Mezzacapo}}\ and\ \bibinfo
  {author} {\bibfnamefont{M.}~\bibnamefont{Boninsegni}},\ }%
  \bibfield{journal}{%
  \bibinfo {journal} {Phys. Rev. Lett.}\ }%
  \textbf{\bibinfo {volume} {97}},\ \bibinfo {pages} {045301} (\bibinfo {year}
  {2006})%
  \bibAnnoteFile{NoStop}{Mezzacapo2006}%
\bibitem{Mezzacapo2007}%
  \BibitemOpen
  \bibfield{author}{%
  \bibinfo {author} {\bibfnamefont{F.}~\bibnamefont{Mezzacapo}}\ and\ \bibinfo
  {author} {\bibfnamefont{M.}~\bibnamefont{Boninsegni}},\ }%
  \bibfield{journal}{%
  \bibinfo {journal} {Phys. Rev. A}\ }%
  \textbf{\bibinfo {volume} {75}},\ \bibinfo {pages} {033201} (\bibinfo {year}
  {2007})%
  \bibAnnoteFile{NoStop}{Mezzacapo2007}%
\bibitem{Cuervo2008}%
  \BibitemOpen
  \bibfield{author}{%
  \bibinfo {author} {\bibfnamefont{J.~E.}\ \bibnamefont{Cuervo}}\ and\ \bibinfo
  {author} {\bibfnamefont{P.-N.}\ \bibnamefont{Roy}},\ }%
  \bibfield{journal}{%
  \bibinfo {journal} {J. Chem. Phys.}\ }%
  \textbf{\bibinfo {volume} {128}} (\bibinfo {year} {2008})%
  \bibAnnoteFile{NoStop}{Cuervo2008}%
\bibitem{Greiner2002}%
  \BibitemOpen
  \bibfield{author}{%
  \bibinfo {author} {\bibfnamefont{M.}~\bibnamefont{Greiner}}, \bibinfo
  {author} {\bibfnamefont{O.}~\bibnamefont{Mandel}}, \bibinfo {author}
  {\bibfnamefont{T.}~\bibnamefont{Esslinger}}, \bibinfo {author}
  {\bibfnamefont{T.~W.}\ \bibnamefont{Hansch}},\ and\ \bibinfo {author}
  {\bibfnamefont{I.}~\bibnamefont{Bloch}},\ }%
  \bibfield{journal}{%
  \bibinfo {journal} {Nature}\ }%
  \textbf{\bibinfo {volume} {415}},\ \bibinfo {pages} {39} (\bibinfo {year}
  {2002})%
  \bibAnnoteFile{NoStop}{Greiner2002}%
\end{thebibliography}
\end{document}